\documentstyle{jpsj2}

\title{%
The performance of thin NaI(Tl) scintillator plate for dark matter search\\
}
\author{%
K.Fushimi$^{\rm{a}}$,
H.Kawasuso$^{\rm{a}}$,
E.Aihara$^{\rm{a}}$,
R.Hayami$^{\rm{a}}$,
M.Toi$^{\rm{a}}$,
K.Yasuda$^{\rm{a}}$,
S.Nakayama$^{\rm{a}}$,
N.Koori$^{\rm{a}}$,
M.Nomachi$^{\rm{b}}$,
K.Ichihara$^{\rm{b}}$,
R.Hazama$^{\rm{b}}$,
S.Yoshida$^{\rm{b}}$,
S.Umehara$^{\rm{b}}$,
K.Imagawa$^{\rm{c}}$,
H.Ito$^{\rm{c}}$
}
\inst{%
a) Faculty of Integrated Arts and Sciences The University of Tokushima\\
1-1 Minami Josanjimacho Tokushima city, 770-8502 Tokushima, Japan\\
b) Department of Physics, Osaka University \\
1-1 Machikaneyamacho Toyonaka city, 570-0043 Osaka, Japan\\
c)Horiba Ltd.\\
2 Kisshoin Miyanohigashimachi Kyoto city, 601-8510 Kyoto, Japan\\
}

\recdate{\today}

\abst{%
A thin (0.05cm) and wide area (5cm$\times$5cm) NaI(Tl) scintillator 
was developed.
The performance of the thin NaI(Tl) plate, energy resolution, single 
photoelectron energy and position sensitivity were tested.
An excellent energy resolution of 20\% (FWHM) at 60keV was obtained.
The single photoelectron energy was calculated to be approximately 
0.42$\pm$0.02keV.\@
Position information in the 5cm$\times$5cm area of the detector was also 
obtained by analyzing the ratio of the number of photons collected 
at opposite ends of the detector.
The position resolution was obtained to be 1cm (FWHM) in the 5cm$\times$5cm 
area.
}

\kword{%
Thin NaI(Tl) scintillator, Energy resolution, Position sensitivity
}

\begin{document}
\maketitle

\section{Introduction}
The need for good position resolution is increasing in the fields 
of nuclear and particle physics.
Because the signals of dark matter and various rare nuclear and particle
processes have the characteristics of a time and space distribution,
position information helps us to distinguish signal events 
from background events.

In the present work, a thin plate of NaI(Tl) scintillator 
has been developed to search for particle candidates of WIMPs 
(Weakly Interacting Massive Particles) dark matter.
WIMPs dark matter has been proposed by both theoretical particle 
physics (SUSY neutralino) and cosmological observation (Cold Dark 
Matter)\cite{Review}.
The search for WIMPs as dark matter candidates is 
some of the most important work in the fields of particle physics
and cosmological physics.

The NaI(Tl) scintillator is suitable for the search for WIMPs
because of the following characteristics.
\begin{itemize}
\item {\bf $^{23}$Na}: 100\% abundance of odd-spin nuclei, large
spin-matrix element $\lambda^{2}J(J+1)=0.089$
for elastic scattering\cite{Ressell}, where $J$ is the total angular 
momentum of $^{23}$Na.
\item {\bf $^{127}$I}: 100\% abundance of odd-spin nuclei. 
The large spin-matrix element for both elastic scattering 
$\lambda^{2}J(J+1)=0.126$
and inelastic scattering $\left|\frac{M_{M1}}{\mu_{p}}\right|^{2}=0.0128$
\cite{Ressell,Ellis},
where $M_{M1}$ is the transition matrix element 
of the M1 transition between the ground state and the first excited state 
of $^{127}$I, and $\mu_{p}$ is the magnetic moment of a proton.
\end{itemize}

In the case of background events, they have the characteristics of 
a space and time distribution.
The most serious background events that hinder the WIMPs signal are U-chain, 
Th-chain and $^{40}$K in the detector.
In the cases of the U-chain and Th-chain, the decays of the progeny 
follow the same position in the characteristic timing.
For example,
the 90\% of $\beta$ rays are followed by the $\alpha$ ray within 
1ms at the same position for the decay chain of 
$^{214}$Bi$\rightarrow^{214}$Po$\rightarrow^{210}$Pb chain. 
$^{210}$Pb is one of the most difficult background source to reduce
\cite{fushimi}.
The low energy $\beta$ and $\gamma$ rays are
followed by the high-energy $\beta$ rays of the long-life progeny of $^{210}$Bi
($T_{1/2}=5.01$day\cite{TOI}).
Because of the long half-life of $^{210}$Bi, an accidental event due to 
the other $^{210}$Pb or $^{210}$Bi occurs between the decay of the parent 
$^{210}$Pb and the decay of its progeny $^{210}$Bi.

Many groups searching for WIMPs have applied a large-volume 
and highly sensitive detector because
the expected cross section of WIMPs-nucleus elastic scattering is
less than 10$^{-6}$pb.
However, it is difficult to observe the precise position where the 
event occurs.
Recently, position-sensitive detectors for WIMPs search have been
proposed by some groups.
The gas detector $\mu-$PIC has been developed to observe the fine tracks 
of charged particles \cite{Miuchi}.
A liquid Xe detector has also been developed for WIMPs search \cite{XMASS}.
A good energy and position resolution was obtained using a
large-volume liquid Xe detector.

We proposed the application of a highly segmented NaI(Tl) detector system to 
search for dark matter\cite{fushimi}.
The sensitivity of this system to WIMPs is largely enhanced by 
investigating the position information of the events.
The segmentation of the detector is shown to be the best way to 
enhance the position sensitivity.
Coincidence measurements of nuclear recoils and $\gamma$ rays for the 
inelastic excitation of $^{127}$I are performed using 
the highly segmented NaI(Tl) detector.
In the case of elastic scattering, the events are usually observed as 
isolated events in space and time.
Recently, ionized atomic electrons and hard X rays following 
WIMPs-nuclear interactions have been shown to be useful for 
the exclusive measurement of
nuclear recoils from the elastic scatterings of WIMPs off nuclei
\cite{ejiri,ejiri2}.
On the other hand, the background events have their own characteristics of 
timing and spatial profiles.
Because the event rate due to the background is reduced by segmentation, 
there is a probability of the accidental coincidence of 
individual background events.
Consequently, the background events are reduced efficiently.
The basic concept of a highly segmented NaI(Tl) array was described
previously\cite{fushimi}.
In this paper, we show the excellent performance of a thin (0.05cm)
and wide-area (5cm$\times$5cm) NaI(Tl) scintillator.

\section{Design of thin NaI(Tl) plate}
The present detector system was developed to search for particle 
candidates of WIMPs dark matter.
Because good position resolution and large acceptance were needed for 
the present project, a thin and wide-area detector was designed.

The design of the thin NaI(Tl) plate is 0.05cm in thickness,
for good position resolution, and 5cm$\times$5cm in 
area for large acceptance.
We aimed to search for WIMPs by coincidence measurements 
of nuclear recoils and the $\gamma$ rays by piling many plates of thin NaI(Tl).
The range of the recoil nucleus is so short that all the energy is 
deposited in one NaI(Tl) plate.
On the other hand, gamma and X rays can escape out of the 
NaI(Tl) plate and they are detected by adjacent NaI(Tl) plates.
In the case of the nuclear excitation of $^{127}$I, a 57.6keV gamma ray is 
emitted from the NaI(Tl) plate.
More than 20\% of the $\gamma$ rays are detected by the adjacent NaI(Tl)
palte when the thickness of the NaI(Tl) plate is 0.05cm.a
If the thickness of the NaI(Tl) plate is less than 0.03cm, 
the $\gamma$ rays that escaped are not detected by the adjacent NaI(Tl) plate.
The gamma ray is detected by the other plates of NaI(Tl) and 
it is difficult to distinguish this from high-energy background events.
We determined the thickness of the NaI(Tl) plate as 0.05cm by 
a Monte Carlo simulation (see ref.\cite{fushimi} for details of the 
determination of thickness).
\begin{figure}[ht]
\includegraphics[width=15cm]{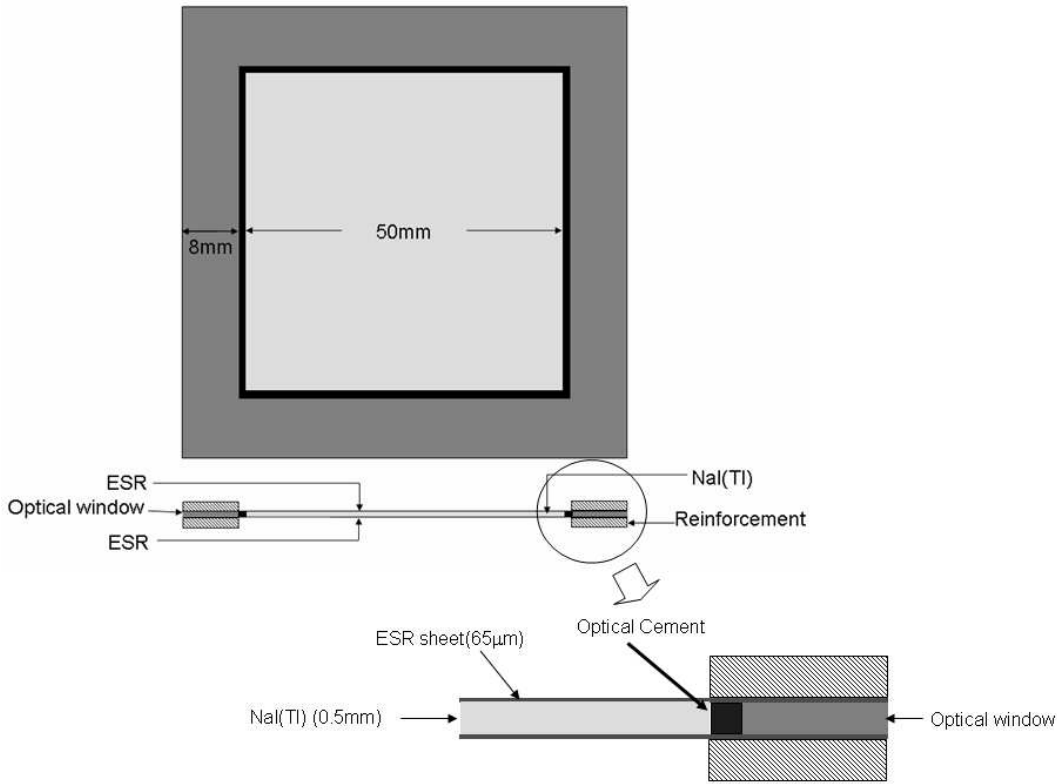}
\caption{
Schematic drawings of thin NaI(Tl) detector.
Upper: Top view of detector. 
Middle: Side view of detector.
Bottom:Enlargement of side view at edge.
}
\label{fg:thinnai}
\end{figure}

We also need a wide-area NaI(Tl) plate, 
because a large acceptance for WIMPs is needed.
Several tens of kg of fiducial mass is needed for WIMPs search 
because of a small cross section ($\sigma< 10^{-5}$pb).
A large-volume detector system was constructed by piling up the wide area 
of NaI(Tl) plates.
We chose an area for the NaI(Tl) plates that was suitable to ensure the
successful production of a large number of thin and wide NaI(Tl) plates.
The schematic of a thin NaI(Tl) plate is shown in 
Fig.\ref{fg:thinnai}.

A reflector with high reflection and low contamination with radioactive 
isotopes is also needed.
Because we need to measure the low energy threshold (a few keV),
there is a need to guide the scintillation photons to the thinner 
edges of the NaI(Tl) crystal.
To guide the photons efficiently to the edges, 
the entire surface of the NaI(Tl) crystal was polished.
The scintillation photon was guided to the edges by total reflection.
An enhanced specular reflector (ESR) was selected to guide 
the photons that failed to go to the edges of the crystal through 
total reflection.
The ESR, which was provided by 3M, was made of polyester whose 
thickness is 65$\mu$m\cite{3M}.
The reflective index for scintillation photons at the ESR is larger than 98\%,
whereas the index at the PTFE sheet is approximately 91\%.
Consequently, a good collection efficiency of scintillation photons 
is expected.
The radioactive contamination of an ESR sheet was also investigated.
The gamma rays from radioactive contamination in the ESR sheet were measured 
by a highly sensitive Ge detector and no contamination beyond the 
background was detected.

Four modules of the thin NaI(Tl) scintillator have been produced by 
Horiba Ltd.
At first, the production of such a thin NaI(Tl) scintillator 
was expected to be difficult because NaI(Tl) is very fragile.
However, we developed a good method of producing thin plates of 
NaI(Tl).
A small crystal of dimensions 5cm$\times$5cm in largest area and 
a few millimeters in thickness was cut out of a large ingot of a NaI(Tl) 
crystal.
One side of the largest surface and four thinner 
surfaces were carefully polished.
The crystal was held on a vacuum chuck and ground carefully, and produced 
a thin NaI(Tl) crystal of thickness 0.05cm.

\section{Energy resolution and energy threshold}
\begin{figure}[ht]
\includegraphics[width=13cm]{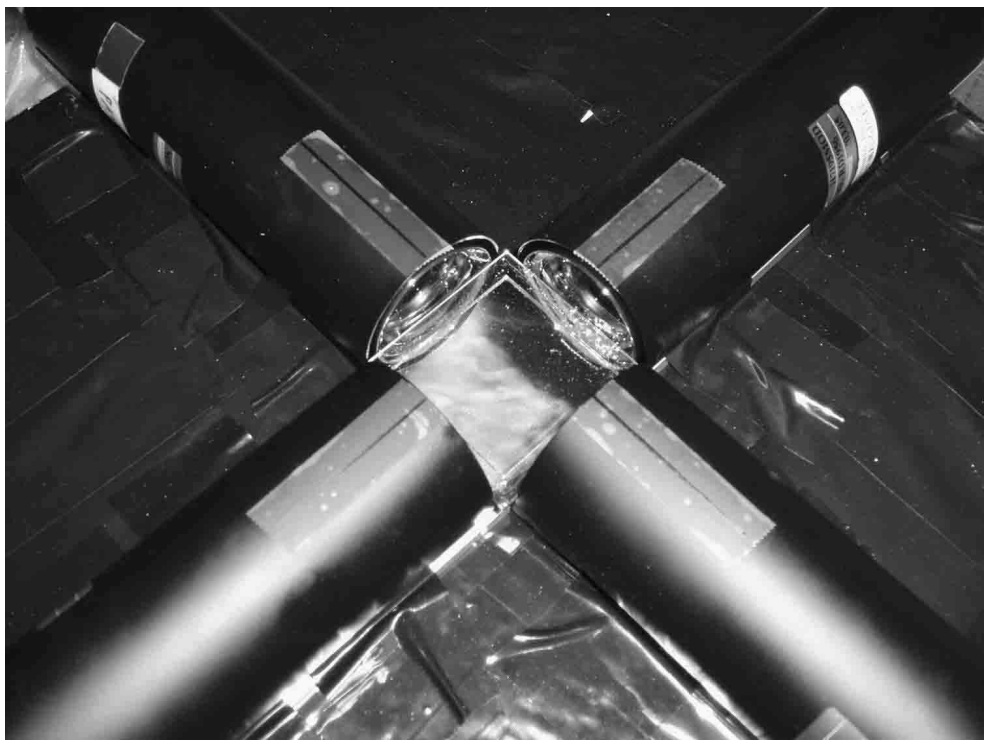}
\caption{
Apparatus for measurement of low energy signals.
Four PMTs were coupled to four edges of the thin NaI(Tl) scintillator.
}
\label{fg:exp}
\end{figure}

\begin{figure}[ht]
\includegraphics[width=16cm]{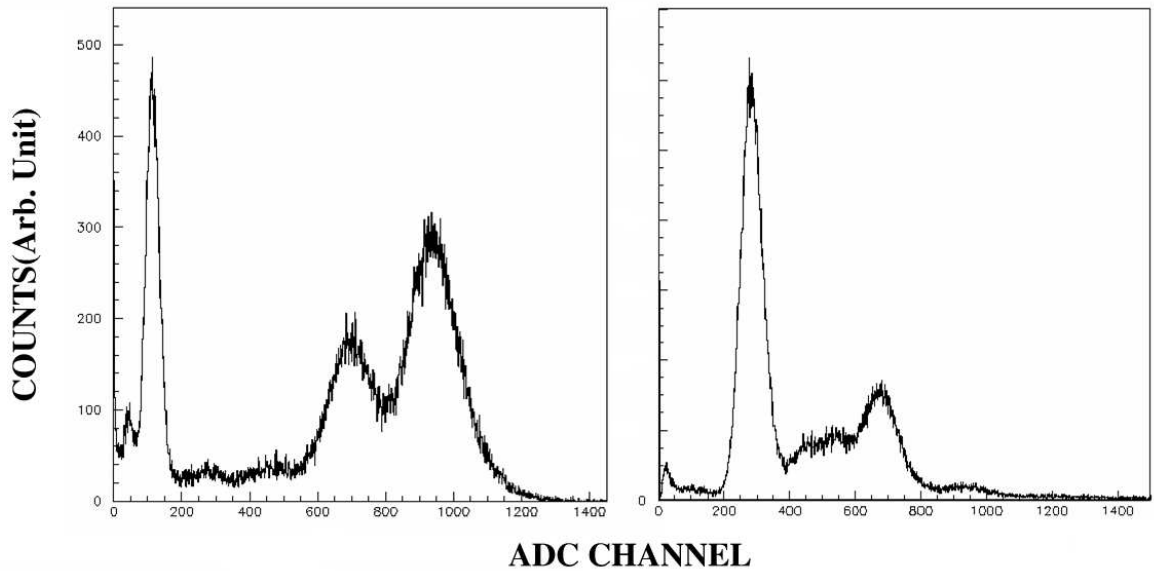}
\caption{
Pulse height spectra of $^{57}$Co (left) and 
$^{133}$Ba (right).
The spectra were obtained using a thin NaI(Tl) scintillator 
and four PMTs.
}
\label{fg:spect}
\end{figure}

The performance of the thin NaI(Tl) scintillator 
was measured by irradiating it with low-energy $\gamma$ rays and X rays.
The scintillation photons were collected at the four edges of 
the NaI(Tl) crystal using four photomultiplier tubes (PMTs),
which were provided by Hamamatsu Photonics (R329-P).
An optical window made of quartz was glued on the thinner edge of the 
NaI(Tl) crystal.
The NaI(Tl) crystal was viewed using the four PMTs at the thinner edges of the 
optical windows.
An image of the apparatus is shown in Fig.\ref{fg:exp}.
The gamma rays and X rays were irradiated isotropically on the 
wider surface of the NaI(Tl) plate.
Single-photoelectron signals were clearly observed using each PMT
and the event rate range was 1$\sim$2kHz.
However, the single photoelectron events were ignored
as they were mainly due to dark current.
Each PMT output signal was individually input into four discriminators.
The threshold of the discriminators was set above 
that of the single-photoelectron signal;
the corresponding hardware energy threshold was 0.8keV.\@
The four PMT outputs were individually converted to digital data using 
a charge integrating analog-to-digital converter (RPC-022).
The gain of each PMT was adjusted by setting the optimum high-voltage.
The total charge outputs of the PMTs were summed event-by-event 
using an off-line analyzer.
The resulting photon outputs are shown in Fig.\ref{fg:spect}.

In the pulse-height spectrum of $^{57}$Co, three prominent peaks and 
one small peak were observed.
The prominent peaks at approximately 950ch and 130ch were photoelectric peaks 
of 122keV and 14.4keV gamma rays.
Another prominent peak at approximately 700ch was due to the photoelectric 
effect of 122keV gamma ray followed by the escape of K-X rays of iodine.
The NaI(Tl) plate was sufficiently thin to allow the escape of a 
large fraction of low-energy X rays.
In the low-energy region of the spectrum, a small but clear 
peak of 6.4keV X rays from iron was observed at approximately 50ch.
This energy spectrum was well reproduced by the Monte Carlo simulation 
using GEANT 4\cite{Geant4}.

In the spectrum of $^{133}$Ba, high-energy gamma rays of energies  
above 200keV were not clearly observed because the detector was too thin to 
absorb the gamma rays.
A photoelectric peak of 81keV and the corresponding X ray escape peak were 
observed at approximately 670ch and 500ch, respectively.
The prominent peak at approximately 300ch was due to the K-X rays of cesium.
Note that the small peak due to 
the low-energy L-X rays of cesium is observed approximately 30ch.
It is relatively important in the search for WIMPs to be able to 
observe low energy, 
and the present results correspond to the energy threshold being 2keV.\@
The results showed that the thin NaI(Tl) scintillator displays  great 
promise in the search for WIMPs.

The energy resolutions at FWHM (Full Width at Half Maximum) were calculated 
from the peaks and are shown in Table \ref{tb:res}.
\begin{table}[ht]
\caption{
Energy resolution for low-energy photons at FWHM.\@
The calculated number of photoelectrons (P.E.) is listed in the fourth column.
}
\label{tb:res}
\hspace*{3.5cm}
\begin{tabular}{lrrr} \hline
Source   & Energy (keV) & $\Delta E/E$ (FWHM) & \# of P.E.\\ \hline
$^{133}$Ba  & 81  & 0.17 & 197\\
$^{241}$Am  & 60  & 0.20 & 143\\
$^{133}$Ba  & 31  & 0.28 & 71\\ 
$^{57}$Co   & 14.4 & 0.40 & 35\\ \hline
\end{tabular}
\end{table}
The collected photoelectron number is directly related to 
the energy resolution by 
\begin{equation}
\frac{\Delta E}{E}=2.354F\frac{\sigma_{N}}{N},
\end{equation}
where $N$ is the collected photoelectron number, 
$\sigma_{N}$ is the standard deviation of the photoelectron number $N$ 
and $F$ is the Fano factor\cite{Knoll}.
Because the collected photoelectron number has a Poisson distribution,
the statistical fluctuation should be $\sigma_{N}=\sqrt{N}$.
The Fano factor, $F=1$ for the scintillators \cite{Knoll}.
Thus, the photoelectron number is derived from 
\begin{equation}
N=\left\{ \frac{2.354E}{\Delta E}\right\}^{2}.
\end{equation}

The scintillation output has good linearity up to 120keV.\@
From the pulse height spectra, the low energy threshold was found to be
2$\sim$3keV.\@
The energy equivalent to a single photoelectron was also 
calculated using the photoelectron number $N$.
The energy threshold of approximately 2$\sim3$keV corresponds to 4$\sim$5 
photoelectrons.
The results showed an excellent performance that is in accordance with 
the required performance for the advanced stage of the experiment.

\section{Position determination}
The position resolution for the thinner directions is as good as 0.05cm 
because of the segmentation of the detector.
Moreover, the position resolution for other directions was tested.
Because the largest area has dimensions of 5cm$\times$5cm,
good position information in the wider area enhances detector sensitivity.
Position information was obtained by analyzing the ratio of 
the number of photons collected on the opposite sides of the detector.
Precise position information on the largest area of the thin NaI(Tl) 
scintillator is important to ascertain the property of the events.
By piling up the thin NaI(Tl) scintillator, the tracking of radiation 
such as cosmic rays and the multiple Compton scattering of photons 
is reconstructed precisely.

\begin{figure}[ht]
\includegraphics[width=13cm]{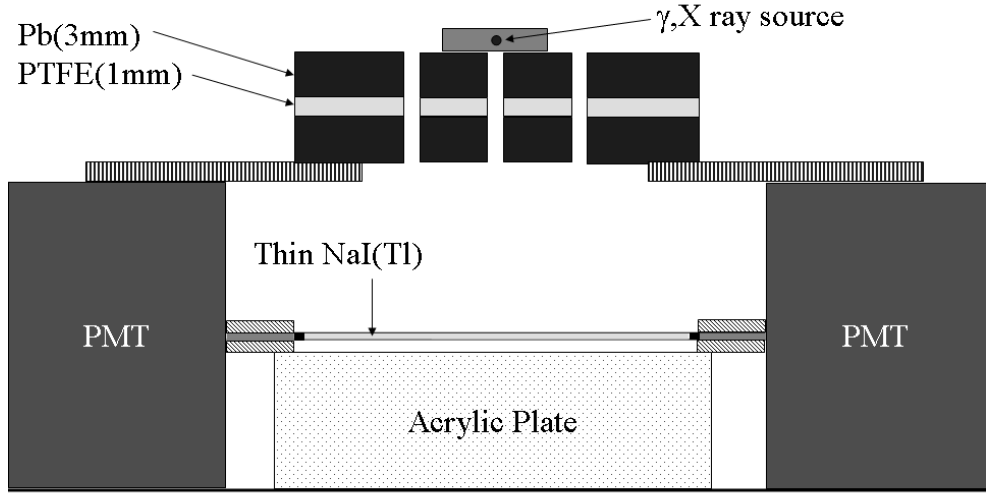}
\caption{
Schematic drawing of cross-sectional view of experiment for 
position resolution in wider area.
}
\label{fg:pos-exp}
\end{figure}

\begin{figure}[ht]
\includegraphics[width=15cm]{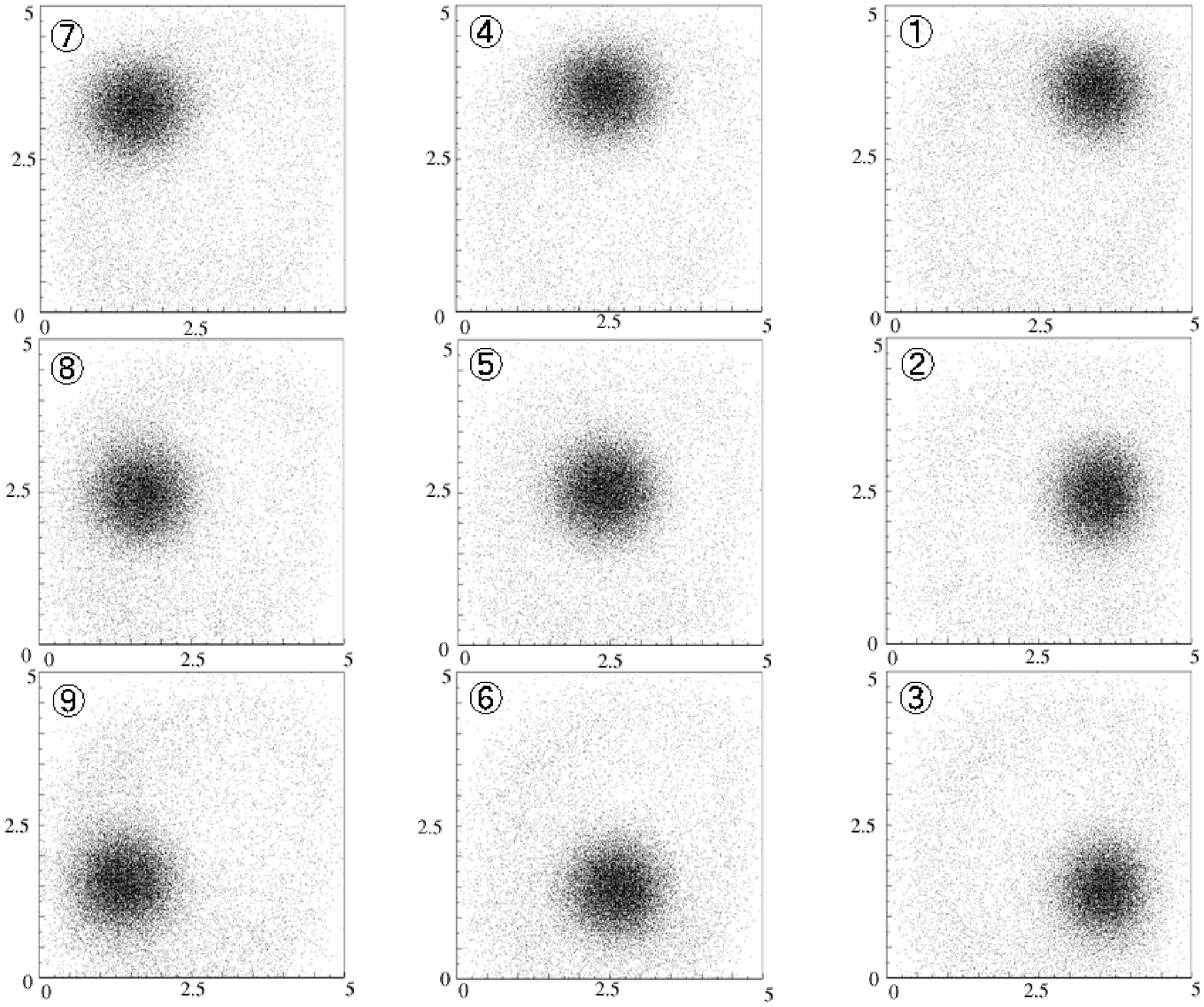}
\caption{
The position analysis result. The $x$ and $y$ axes correspond to the 
real position in NaI(Tl) crystal in unit of cm.
}
\label{fg:posi}
\end{figure}

A collimator for low-energy $\gamma$ rays was made of 1cm-thick lead brick.
Nine holes with a diameter of 2mm were drilled through the lead brick.
An $^{241}$Am source was placed at the top of 
each hole(see Fig.\ref{fg:pos-exp}).
Position determination analysis was performed using 60keV gamma 
rays from $^{241}$Am.
The charge outputs from each PMT are defined as 
$C_{i}, i=1,2,3,4$.
The position where a $\gamma$ ray was observed was calculated by 
\begin{eqnarray}
x & = & a+b\frac{C_{3}-C_{1}}{C_{3}+C_{1}} \\ \nonumber
y & = & c+d\frac{C_{4}-C_{2}}{C_{4}+C_{2}} , \nonumber
\end{eqnarray}
where $a,b,c$ and $d$ are the parameters to normalize the positions $x$ and 
$y$ as the real positions. 
The results of the position analysis are shown in Fig.\ref{fg:posi}.
The position resolution was calculated to be approximately 1cm FWHM.\@

Further segmentation in the wide area of the thin NaI(Tl) plate
is achieved by position analysis.
The background events due to $^{210}$Pb are effectively reduced by the
segmentation.
The number of scintillation photons for 46.5keV $\gamma$ rays 
of $^{210}$Pb is approximately 3/4 of that for the 60keV $\gamma$ ray of 
$^{241}$Am; however, the position resolution degrades by only 10\%.
The FWHM of the position accuracy was approximately 1cm; thus, 
the 3$\times$3 segmentation 
in the wide area of 5cm$\times$5cm was certainly achieved.

\section{Conclusion and Discussion}
We showed that measuring the precise position of 
the interaction of radiation enhances sensitivity in nuclear- and 
particle-rare physics.
The optical segmentation of the detector achieves extreme position 
sensitivity.
We successfully developed a thin (0.05cm in thickness) and wide-area 
(5cm$\times$5cm) NaI(Tl) scintillator plate.

We tested the performance of this thin NaI(Tl) plate and obtained  
good performance.
The energy threshold was directly measured as approximately 2keV and the 
single-photoelectron energy was obtained as 0.42$\pm$0.02keV.\@
The position sensitivity was 0.05cm in one direction because of the 
optical segmentation.
Moreover, approximately 1cm in FWHM of position resolution 
in the wider area was obtained by the roll-off analysis.

The enhancement in sensitivity due to position selectivity 
will be discussed next.
A highly pure NaI(Tl) with a $^{210}$Pb contamination as small as 0.1mBq/kg  
is now under development.
It is quite difficult to reduce the $^{210}$Pb concentration because
$^{210}$Pb is contaminated by small amounts of normal air and water.
In the case of the NaI(Tl) crystal whose contamination is 0.1mBq/kg of 
$^{210}$Pb, the event rate due to $^{210}$Pb is 0.0397/day in one 
plate of the thin NaI(Tl).
The time-correlation analysis is performed because the accidental event rate 
of $^{210}$Pb is fivefold longer than the half-life of $^{210}$Bi (5.01days).
Practically, it is effective to remove the event if the following event 
occurs within 12.5days and the background rate is reduced by a factor of 
$8.2\times 10^{-2}$.
By applying position analysis, the event rate in one segment is reduced 
ninefold the magnitude and the event rate of $^{210}$Pb is expected to be
$4.4\times 10^{-3}$/day.
Note that the individual $^{210}$Pb decay does not occur within
 more than 220days in the same NaI(Tl) plate.

In the case of high-energy charged particles, they deposit more than 
300keV in the NaI(Tl) plate.
Position resolution is enhanced by the factor 
$\left(\frac{60{\rm keV}}{300{\rm keV}}\right)^{1/2}\simeq0.5$, thus, the 
position resolution in the wider area is 0.5cm.
The array of the thin NaI(Tl) scintillator is expected to be highly effective
for particle tracking, even in a low-energy region.

\section{Acknowledgment}
The authors thank Professors T.Kishimoto and M.Minowa for fruitful 
discussions and encouragement.
The authors also thank Dr.\ Y.\ Yanagida and Mr.\ M.\ Hori 
for discussions 
and making the first version of the detector plate.
This work was supported by Toray Science Foundation and in part by 
The Faculty Grant of Integrated Arts and Sciences, Tokushima University.

\end{document}